\documentclass[11pt]{article}
\usepackage{amsfonts}
\usepackage{amsmath,amssymb}

\newcommand{\hko}{\hookrightarrow}

\newcommand{\n}{\nonumber\\}

\newcommand{\bec}{\begin{center}}
\newcommand{\eec}{\end{center}}

\newcommand{\bea}{\begin{array}}
\newcommand{\ear}{\end{array}}

\newcommand{\bfr}{\begin{flushright}}

\newcommand{\efr}{\end{flushright}}
\newcommand{\noi}{\noindent}
\newcommand{\me}{\frac{1}{2}}

\newcommand{\cl}{{\mt{C}}\ell}
\newcommand{\RR}{\mathbb{R}}\newcommand{\op}{\oplus}
\newcommand{\HH}{\mathbb{H}}\newcommand{\PP}{\mathbb{P}}
\newcommand{\ap}{\alpha}

\newcommand{\ot}{\otimes}
\newcommand{\la}{\Lambda}
\newcommand{\bege}{\begin{equation}}
\newcommand{\enge}{\end{equation}}
\newcommand{\w}{\wedge}
\newcommand{\g}{\gamma}
\newcommand{\om}{\omega}
\newcommand{\ri}{\rightarrow}

\newcommand{\ty}{\RR\oplus\RR^3}
\newcommand{\si}{\sigma}

\newcommand{\beq}{\begin{eqnarray}}\newcommand{\benu}{\begin{enumerate}}\newcommand{\enu}{\end{enumerate}}
\newcommand{\eeq}{\end{eqnarray}}
\newcommand{\mt}{\mathcal}

\newcommand{\vv}{{\bf v}}
\newcommand{\ee}{{\bf e}}
\newcommand{\uu}{{\bf u}}

\newcommand{\ff}{{\bf f}}
\newcommand{\ww}{{\bf w}}
\newcommand{\CC}{\mathbb{C}}

\newcommand{\ZZ}{\mathbb{Z}}
\newcommand{\ep}{\epsilon}
\newcommand{\bmw}{\mathbf{w}}

\newcommand{\ol}{\overline}
\newcommand{\vbn}{\blacktriangleleft}
\newcommand{\vvn}{\blacktriangleright}

\newcommand{\xx}{{\bf x}}

\newcommand{\clt}{{\mt{C}}\ell_{3,0}}
\newcommand{\cle}{{\mt{C}}\ell_{1,3}}

\newcommand{\BA}{\breve{A}}

\newcommand{\BB}{\breve{B}}

\newcommand{\bx}{\begin{pmatrix}}
\newcommand{\ex}{\end{pmatrix}}
\newcommand{\vcx}{\varepsilon}
\newcommand{\mma}{{\mathfrak{a}}}
\newcommand{\mmb}{{\mathfrak{b}}}
\newcommand{\mmg}{{\mathfrak{g}}}

\usepackage{amsfonts}
\usepackage{amsmath,amssymb}

\usepackage{indentfirst}  
\begin{document}
\title{Revisiting Clifford algebras and spinors III: conformal structures and twistors in the paravector model of  spacetime}
\author{{\bf Rold\~ao da Rocha}\thanks{Instituto de F\'{\i}sica Gleb Wataghin (IFGW), Unicamp, CP 6165, 13083-970, Campinas (SP), Brazil. 
E-mail: roldao@ifi.unicamp.br. Supported by CAPES.}\and{\bf Jayme Vaz, Jr.}\thanks{Departamento de Matem\'atica Aplicada, IMECC, Unicamp, CP 6065, 13083-859, Campinas (SP), Brazil. E-mail: 
vaz@ime.unicamp.br}}
\date{}\maketitle

\abstract${}^{}$\begin{center}
\begin{minipage}{12cm}$ \;\;\;\;\;$
This paper is the third of a series of three, and it is the continuation of 
\cite{rol1,rol2}. After reviewing the conformal spacetime structure,  
 conformal maps are described  in Minkowski spacetime as the twisted adjoint representation of  \$pin$_+$(2,4), acting  
 on paravectors. Twistors are then presented via the paravector model of Clifford algebras
and related to conformal maps in the Clifford algebra over the lorentzian $\RR^{4,1}$ spacetime. We construct
 twistors in Minkowski spacetime as algebraic spinors associated with the Dirac-Clifford algebra
 $\mathbb{C}\otimes C\ell_{1,3}$ 
 using one lower spacetime dimension than standard Clifford algebra formulations, since for this purpose 
 the Clifford algebra over $\RR^{4,1}$ is also used to describe conformal maps, instead of $\RR^{2,4}$.
Although some papers have already described twistors using the algebra $\CC\otimes\cl_{1,3} \simeq \cl_{4,1}$, the present formulation
sheds some new light on the use of the paravector model and generalizations.

\end{minipage}\end{center}
\medbreak
\medbreak\noi\newpage
\section*{Introduction}

Twistor theory is originally based on spinors, from the construction of a space, the {\it twistor space}, 
in such a way that the spacetime structure emerges as a secondary concept. 
According to this formalism, twistors are considered as more primitive entities than spacetime points.
Twistors are used to describe some physical concepts, for example,  {momentum}, angular momentum, helicity
and massless fields \cite{Be93,Be96}. The difficulties to construct a theory for quantum gravity, based on the continue spacetime structure,
suggests the discretization process of such structure \cite{Co94}. 
One of the motivations to investigate twistor theory are the {\it spin networks}, related to a discrete description of spacetime \cite{pe3,pe4,pe5}. 

Twistor formalism has been used to describe a lot of physical theories, 
and an  increasing progress of wide-ranging applications of this formalism, via Clifford algebras, has been done \cite{BH85,Kl74,Ko96,AO82,Cw91} in the last two decades. 
 Another branch of applications is the union between twistors, supersymmetric theories and strings (see, for example \cite{Ho95,LD92,Bk91a,Bk91b,Wi86a,mot,b1,b2,b4,grscwi}
and many others). 
As a particular case, the classical Penrose twistor formalism \cite{pe3,pe4,pe5,pe1,pe2} describes a  spin 3/2 particle, the gravitino, which is
the graviton superpartner. 
Twistor formalism is also used in the investigation about the relativistic dynamics of elementary particles \cite{Be00} and 
 about confined states  \cite{KA96}.
 
The main aim of the present paper is to describe spinors and twistors as algebraic objects, from the Clifford algebra standview. 
In this approach, a twistor is an {algebraic spinor} \cite{chev}, an element of a lateral minimal ideal of a Clifford algebra. This
characterization is done using the representation of the conformal group and the structure of the Periodicity Theorem of Clifford algebras \cite{ABS,benn,port,coq}.
Equivalently, a twistor is an element that carries the representation of the group $\${\rm pin}_+(2,4)$, 
the double covering\footnote{The double covering of SO$_+$(2,4) is the group Spin$_+$(2,4). The notation \$pin$_+$(2,4) 
is motivated by the use of paravectors, elements of  $\RR\op\RR^{4,1}\hko\cl_{4,1}$.} of SO$_+$(2,4).

The group SO$_+$(2,4), that describes proper orthochronous rotations in $\RR^{2,4}$, 
is the invariance group of the bilinear invariants \cite{Lo96} in the Dirac relativistic quantum mechanics theory \cite{itz}. This group 
is also the double covering of SConf$_+$(1,3), the group of the proper special conformal transformations, that is the 
biggest one that preserves the structure of Maxwell equations, leaving invariant the light-cone in Minkowski spacetime. 

We could cite other, more applicated, motivations to investigate twistors. For instance, a twistor is also written  as the sum of a
 light-like vector with the dual of a vector $\xx\in\RR^{1,3}$, multiplied by the same light-like vector \cite{ke97}.
In this way, it describes the structure of a vector $\xx$ that supports another light-like vector \cite{pe3,BH85,pe1}. 
Such structures are still generalized, if {\it screws} and {\it mexors} are defined \cite{ke97}. The screws describe analogously 
a vector supporting another vector, but now the condition asserting that  the two vectors must be light-like does not need
to be satisfied. Robots moviments, homocinetic articulations and conjugate arms are still investigated, using screws and mexors.  
The formalism related are mathematically easy, presented at an undergraduate level \cite{Gu97,Ku99} and useful to important applications, 
since they are associated, for example, to the study of human eyes moviments  \cite{He94a} and
 neurogeometry \cite{Pl80,TC90,He94b}. Mexors are geometric entities describing eletrons with spin and their interactions \cite{ke97}.

 This paper is presented as follows: in Sec. 1 we present some mathematical preliminaries and fix the notation to be used. 
In Sec. 2 we first enunciate and prove the Periodicity Theorem of Clifford algebras, from which and M\"obius maps in the plane
are investigated. We also introduce the conformal compactification of $\RR^{p,q}$ and then the conformal group is defined.
In Sec. 3 the conformal transformations in MInkowski spacetime are presented as  
the twisted adjoint representation of the group SU(2,2) $\simeq\$$pin$_+$(2,4) on paravectors of $\cl_{4,1}$. Also, 
the Lie algebra of associated groups and the one of the conformal group, are presented. 
In Sec. 4 twistors, the incidence relation between twistors and the Robinson congruence,
 via multivectors and the paravector model of $\CC\ot\cle\simeq\cl_{4,1}$, are introduced. We show explicitly how our
 results can be led to the well-established ones of Keller \cite{ke97},
and consequently to the classical formulation introduced by Penrose \cite{pe1,pe2}. \\

\section{Preliminaries}
Let $V$ be a finite $n$-dimensional real vector space. We consider the tensor algebra $\bigoplus_{i=0}^\infty T^i(V)$ from which we
restrict our attention to the space $\Lambda(V) = \bigoplus_{k=0}^n\Lambda^k(V)$ of multivectors over $V$. $\Lambda^k(V)$
denotes the space of the antisymmetric
 $k$-tensors, the $k$-forms.  Given $\psi\in\Lambda(V)$, $\tilde\psi$ denotes the \emph{reversion}, 
 an algebra antiautomorphism
 given by $\tilde{\psi} = (-1)^{[k/2]}\psi$ ([$k$] denotes the integer part of $k$). $\hat\psi$ denotes 
the \emph{main automorphism or graded involution}, given by 
$\hat{\psi} = (-1)^k \psi$. The \emph{conjugation} is defined by the reversion followed by the main automorphism.
  If $V$ is endowed with a nondegenerate, symmetric, bilinear map $g: V\times V \rightarrow \RR$, it is 
possible to extend $g$ to $\la(V)$. Given $\psi=u_1\w\cdots\w u_k$ and $\phi=v_1\w\cdots\w v_l$, $u_i, v_j\in V$, one defines $g(\psi,\phi)
 = \det(g(u_i,v_j))$ if $k=l$ and $g(\psi,\phi)=0$ if $k\neq l$. Finally, the projection of a multivector $\psi= \psi_0 + \psi_1 + \cdots + \psi_n$,
 $\psi_k \in \la^k(V)$, on its $p$-vector part is given by $\langle\psi\rangle_p$ = $\psi_p$. 
The Clifford product between $v\in V$ and $\psi\in\la(V)$ is given by $v\psi = v\w \psi + v\cdot \psi$.
 The Grassmann algebra $(\la(V),g)$ 
endowed with this product is denoted by $\cl(V,g)$ or $\cl_{p,q}$, the Clifford algebra associated to $V\simeq \RR^{p,q},\; p + q = n$.

\section{Periodicity Theorem, M\"obius maps and the conformal group}
The Periodicity Theorem if Clifford algebras has great importance and shall be used in the rest 
of the paper. Then it is natural to enunciate and prove the\\\medbreak
{\bf{Periodicity Theorem}}
$\vvn$ {\it Let $\mt{C}\ell_{p,q}$ be the Clifford algebra of the quadratic space $\mathbb{R}^{p,q}$.
The following isomorphisms are verified:
\bege\label{per1}\bea{|l|}\hline{\rule[-3mm]{0mm}{8mm} 
{\mt{C}\ell}_{p+1,q+1}  \simeq {\mt{C} \ell}_{1,1}\otimes {\mt{C}\ell}_{p,q},}\\ \hline\ear
\enge
\bege\label{per2}
{\mt{C}\ell}_{q+2, p}  \simeq {\mt{C} \ell}_{2,0}\otimes{\mt{C}\ell}_{p,q},
\enge
\bege\label{per3}
{\mt{C}\ell}_{q,p+2}  \simeq {\mt{C}\ell}_{0,2} \otimes{\mt{C}\ell}_{p,q},
\enge
where} $p > 0$ or $q > 0$.  $\vbn$
\medbreak

\noi {\textbf{Proof}}: Let $V$ be a 2-dimensional space, 
where  the bilinear symmetric non-degenerate 2-form $g_V: V \times V \rightarrow \RR$ is defined.
In terms of an  orthonormal basis $\{\ff_1,\ff_2\}$,
we have, for $\uu = u^1\ee_1 + u^2 \ee_2 \in V$ the expression
\bege
g_{V} = \lambda_1 (u^1)^2 + \lambda_2 (u^2)^2,
\enge\noi 
where $\lambda_1$ and $\lambda_2$ are chosen to be $\lambda_1 = \pm 1, \lambda_2 = \pm 1,$
 depending on the considered case: $V = \mathbb{R}^{2,0}$, $\mathbb{R}^{1,1}$
or $\mathbb{R}^{0,2}$. The linear application
 $\Gamma : \mathbb{R}^{p,q} \oplus V \rightarrow
\mt{C}\ell(V,g_{V}) \otimes \cl_{p,q}$ is defined by:
$$
\Gamma(\vv + \uu) = \ff_1
\ff_2\otimes \vv +
\uu \otimes 1 , \qquad \forall \uu \in V,\; \vv \in
\mathbb{R}^{p,q},
$$
\noi where it is implicit that $\ff_1 = \rho(\ff_1)$,
$\ff_2 = \rho(\ff_2)$, $\uu$ denotes $\rho(\uu)$ by abuse of notation, and also
$\vv = \gamma(\vv)$,  where $\rho$ and $\gamma$ are the Clifford applications
 $\rho : V \rightarrow \cl(V,g_{V})$ and  $\gamma : \mathbb{R}^{p,q} \rightarrow \cl_{p,q}$.
 $\Gamma$ is indeed a Clifford application:
\beq
\big(\Gamma(\vv + \uu)\big)^2 &=&
\big(\ff_1\ff_2\otimes \vv + \uu \otimes 1)
\big(\ff_1\ff_2\otimes \vv + \uu \otimes 1) \nonumber\\
&=& (\ff_1\ff_2)^2
\otimes (\vv)^2 +
(\uu\ff_1\ff_2 + \ff_1\ff_2\uu)\otimes 
\vv + (\uu)^2\otimes 1 .\nonumber
\eeq
Since $\{\ff_1,\ff_2\}$
is orthogonal, i.e.,  $\ff_1\ff_2 = \ff_1\wedge\ff_2 =
-\ff_2\ff_1$ we have
$$
\uu \ff_1\ff_2 + \ff_1\ff_2\uu =
\uu(\ff_1\wedge\ff_2) + (\ff_1\wedge\ff_2)\uu =
2 \uu\wedge\ff_1\wedge\ff_2 = 0,
$$
because $\uu \in V$ is written as $\uu = a \ff_1 + b\ff_2$.
Using
$$
(\ff_1\ff_2)^2 =  - (\ff_1)^2 (\ff_2)^2 = -\lambda_1\lambda_2 ,
$$
it follows that
\beq
\big(\Gamma(\vv + \uu)\big)^2 &=&
-\lambda_1\lambda_2\otimes g(\vv,\vv) + g_{V}(\uu,\uu)
\otimes 1 \nonumber\\
&=& \big[\lambda_1(u^1)^2 + \lambda_2(u^2)^2 -
\lambda_1\lambda_2 [ (v^1)^2 + \cdots + (v^p)^2 -
(v^{p+1})^2 - \cdots - (v^n)^2]\big] 1\otimes 1.\nonumber
\eeq
So we proved that  $\Gamma$ is indeed a 
Clifford application $\Gamma : \mathbb{R}^{p,q}\oplus V \rightarrow
\cl(W,g_{W})$, where $W$ is a  $(n+2)$-dimensional
vector space, endowed with a bilinear symmetric non-degenerate 2-form $g_{W}$
given by
$$
g_{W}(\bmw,\bmw) =
\lambda_1(u^1)^2 + \lambda_2(u^2)^2 -
\lambda_1\lambda_2 \big[ (v^1)^2  \quad + \cdots + (v^p)^2 -
(v^{p+1})^2 - \cdots - (v^n)^2\big] ,
$$
where $\ww = u^1\ff_1 + u^2\ff_2 + v^i\ee_i$.

Therefore,
\begin{itemize}\item{$ \;{\rm if} \;V = \mathbb{R}^{2,0} \;{\rm then}\; W =
\mathbb{R}^{q+2,p},$}
\item $\;{\rm if}\; V = \mathbb{R}^{1,1} \; {\rm then}\;
W = \mathbb{R}^{p+1,q+1},$
\item $\;{\rm if}\; V = \mathbb{R}^{0,2}\;
{\rm then}\; W = \mathbb{R}^{p,q+2}.$
\end{itemize}

\noi The isomorphisms follow from the  universality property of the Clifford algebras. 
\begin{flushright}{$\Box$}
\end{flushright}

\noi{\textbf{Observation}}: The isomorphism given by eq.(\ref{per1}), the so-called {\bf Periodicity Theorem} 
\bege\bea{|l|}\hline{\rule[-3mm]{0mm}{8mm}
{\mt{C}\ell}_{p+1,q+1}  \simeq {\mt{C} \ell}_{1,1}\otimes {\mt{C}\ell}_{p,q}}\\ \hline\ear
\enge
\noindent is of primordial importance in what follows, since {\it twistors} are characterized via a representation of the conformal group. 
\medbreak

Now we reproduce the result in \cite{ABS,benn,maks}. The { reversion} is denoted by $\ap_1$, while  the { conjugation}, by 
$\ap_{-1}$, in order to simplify the notation. The two antiautomorphisms are included in the notation ${\ap_\ep} (\ep = \pm 1)$. 
 \medbreak
{\bf{Periodicity Theorem (II)}}
$\vvn$ {\it The Periodicity Theorem (\ref{per1}) is given by}
$$\bea{|l|}\hline{\rule[-3mm]{0mm}{8mm} 
(\cl_{p+1,q+1},\ap_\ep) \simeq (\cl_{p,q},\ap_{-\ep}) \ot (\cl_{1,1},\ap_\ep).}\\ \hline\ear\vbn
$$\medbreak\noi\textbf{Proof}: The bases $\{\ee_i\}$, $\{\ff_j\}$ span the algebras $\cl_{p,q}$ and $\cl_{1,1}$, respectively. 
Let $\{\ee_i \ot \ff_1 \ff_2, 1 \ot \ff_j\}$ be a basis for $\cl_{p+1,q+1} \simeq \cl_{p,q} \ot \cl_{1,1}$. 
The following relations are easily verified:
\beq
 (\ap_{-\ep} \ot \ap_\ep)(\ee_i \ot \ff_1 \ff_2) &=& \ap_{-\ep}(\ee_i) \ot \ap_\ep (\ff_1 \ff_2)\n
 &=& \ep (\ee_i \ot \ff_1 \ff_2)\eeq\noi \beq
 (\ap_{-\ep} \ot \ap_\ep)(1 \ot \ff_j) &=& \ap_{-\ep}(1) \ot \ap_\ep (\ff_j)\n
 &=& \ep (1 \ot \ff_j)\eeq\noi Therefore the generators are multiplied by $\ep$.
\bfr$\Box$\efr

\subsection{M\"obius maps in the plane} 
It is well-known that rotations in the  Riemann sphere $\CC\PP^1$ are associated to rotations in the Argand-Gauss plane \cite{pe2}, which is (as a vector space) isomorphic
to $\RR^2$. The algebra $\cl_{0,1}\simeq\CC$ is suitable to describe rotations in $\RR^2$. 
  
From the periodicity theorem of Clifford algebras
\bege
\boxed{\cl_{p+1,q+1}\simeq\cl_{1,1}\ot\cl_{p,q}}
\enge\noi it can be seen that  Lorentz transformations in spacetime, generated by the vector representation of the group  \$pin$_+$(1,3)$\hko\clt$, 
are directly related to the M\"obius maps in the plane, since we have the following correspondence:
  \bege\label{cvbnm}
  \clt\simeq\cl_{2,0}\ot\cl_{0,1}\simeq \cl_{1,1}\ot\cl_{0,1}.
  \enge\noi In this case  conformal transformations are described using $\cl_{1,1}$. 
  From eq.(\ref{cvbnm}) it is possible to represent a paravector \cite{bay2,bayoo} $\mma\in\clt$ in ${\mt M}(2,\CC)$:
  \bege\label{plu}
   \mma =  \left(\bea{cc}z &\lambda\\
 \mu&{\bar{z}}
 \ear\right)\in\RR\op\RR^3,\enge
 \noi where $z\in \CC$, $\mu,\lambda\in\RR$. 
 
 Consider now an element of the group
\bege\${\rm pin}_+(1,3) := \{\phi \in \cl_{3,0} \;|\;\phi\ol{\phi} = 1\}.
\enge\noi From the Periodicity Theorem (II) it follows that
 \bege\label{sgj}
 {\widetilde{\left(\bea{cc} a&c\\
 b&d
 \ear\right)}} = \left(\bea{cc} {\bar d}&{\bar c}\\
 {\bar b}&{\bar a}
 \ear\right).\enge
 
 The rotation of a paravector $\mma\in\RR\op\RR^{3}$ 
can be performed by the twisted adjoint representation  \bege\label{sgj1}
\mma\mapsto\mma' = \eta\mma{\tilde\eta}, \quad \eta\in\${\rm pin}_+(1,3).
\enge\noi  In terms of the matrix representation we can write eq.(\ref{sgj1}), using eq.(\ref{plu}), as: 
 \bege
 \left(\bea{cc} z&\lambda\\
 \mu&{\bar{z}}
 \ear\right)\mapsto\left(\bea{cc} a&c\\
 b&d
 \ear\right) \left(\bea{cc} z&\lambda\\
 \mu&{\bar{z}}
 \ear\right){\widetilde{\left(\bea{cc} a&c\\
 b&d
 \ear\right),}}\enge\noi and using eq.(\ref{sgj}), it follows that
\bege
\left(\bea{cc} z&\lambda\\
 \mu&{\bar{z}}
 \ear\right)\mapsto \left(\bea{cc} a&c\\
 b&d
 \ear\right) \left(\bea{cc} z&\lambda\\
 \mu&{\bar{z}}
 \ear\right)\left(\bea{cc} {\bar d}&{\bar c}\\
 {\bar b}&{\bar a}
 \ear\right)
            \enge
\noi Taking  $\mu = 1$ and $\lambda = z{\bar{z}}$, we see that the paravector  $\mma$ is maped on 
\bege\label{abcd}
\left(\bea{cc} a&c\\
 b&d
 \ear\right) \left(\bea{cc} z&z{\bar{z}}\\
 1&{\bar{z}}
 \ear\right)\left(\bea{cc} {\bar d}&{\bar c}\\
 {\bar b}&{\bar a}
 \ear\right) = \omega\left(\bea{cc} z'& z'{\bar{z'}}\\
 1&{\bar{z'}}
 \ear\right),
 \enge
\noi where $z' := \frac{az + c}{bz + d}$ and $\omega := |bz + d|^2\in\RR$.
 
The map given by eq.(\ref{abcd}) is the spin-matrix ${\bf A}\in {\rm SL}(2,\CC)$, described in \cite{pe2}.

\subsection{Conformal compactification}
The results in this section are achieved in \cite{port,ort}. Given the quadratic space $\RR^{p,q}$, consider the injective map given by
\beq
\varkappa: \RR^{p,q} &\rightarrow&  \RR^{p+1,q+1}\nonumber\\
                  x&\mapsto& \varkappa(x) = (x, x\cdot x, 1) = (x, \lambda, \mu)
                  \eeq

\noi The image of $\RR^{p,q}$ is a subset of the quadric $Q\hko\RR^{p+1,q+1}$, described by the equation:
\bege\label{klein}
x\cdot x - \lambda\mu = 0,
\enge\noi the so-called {\it Klein absolute}.
\noi The map $\varkappa$ induces an injective map from  $Q$ in the projective space $\RR\PP^{p+1,q+1}$. 
Besides, $Q$ is compact and defined as the conformal compactification ${\widehat{\RR^{p,q}}}$ of $\RR^{p,q}$. 

 $Q \thickapprox {\widehat{\RR^{p,q}}}$ is homeomorphic to $(S^p\times S^q)/\ZZ_2$ \cite{ort}.
In the particular case where $p = 0$ and $q = n$, the quadric is homeomorphic to the $n$-sphere $S^n$, 
the compactification of  $\RR^n$ via the addition of a point at infinity. 

There also exists an injective map
\beq s:\RR\oplus\RR^3 &\ri& \RR\oplus\RR^{4,1}\nonumber\\
         v&\mapsto& s(v) = \left(\bea{ll}v&v\bar{v}\\1&{\bar{v}}\ear\right)\eeq
         \noi 
The following theorem is introduced by Porteous \cite{port,ort}:
\medbreak
{\bf{Theorem}} {\it $\vvn$ {\rm (i)} the map $\varkappa: \RR^{p,q}\ri \RR^{p+1,q+1};\; x\mapsto (x,  x\cdot x, 1)$, is an isometry.\\
${}\hspace{3.2cm} {\rm (ii)}$ the map $\pi:Q \ri \RR^{p,q};\;(x, \lambda, \mu)\mapsto x/\mu$ defined where $\lambda\neq 0$ is conformal.\\
${}\hspace{3.2cm} {\rm (iii)}$ if $U:\RR^{p+1,q+1}\ri \RR^{p+1,q+1}$ is an orthogonal map, the the map
 $\Omega = \pi\circ U \circ \varkappa:\RR^{p,q}\ri \RR^{p,q}$ is conformal.} $\vbn$
\medbreak
\noi The application  $\Omega$ maps  conformal spheres onto conformal spheres, which can be {quasi-spheres} or hiperplanes.
A quasi-sphere is a submanifold of $\RR^{p,q}$, defined by the equation 
\bege
a\;x\cdot x + b\cdot x + c = 0,\quad a, c\in \RR,\;\;b\in\RR^{p,q}.
\enge\noi  A quasi-sphere {\it is} a sphere when a quadratic form
  $g$ in $\RR^{p,q}$ is positive  defined and $a\neq 0$. A quasi-sphere is a plane when $a=0$. 
From the assertion $(iii)$ of the theorem above, we see that $U$ and $-U$ induces the same conformal transformation
in $\RR^{p,q}$. The  {\it conformal group} is defined as 
\bege\label{conft} \bea{|l|}\hline{\rule[-3mm]{0mm}{8mm}{\rm Conf}(p,q) \simeq  {\rm O}(p+1,q+1)/\ZZ_2}\\ \hline\ear
\enge
 ${\rm O}(p+1,q+1)$ has four components and, in the Minkowski spacetime case, where $p=1, q=3$, 
the group Conf(1,3) has four components. 
The component of Conf$(1,3)$ connected to the identity, denoted by Conf$_+(1,3)$ 
is known as {\it the M\"obius group} of $\RR^{1,3}$. Besides, SConf$_+$(1,3) denotes the component connected to the identity, time-preserving and future-pointing.

\section{Paravectors of  $\cl_{4,1}$ in $\clt$ via the Periodicity Theorem}
\label{dcv}
Consider the basis $\{\varepsilon_{\BA}\}_{\BA = 0}^5$ of $\RR^{2,4}$ that obviously satisfies the relations
\bege\vcx_0^2 = \vcx_5^2 = 1, \quad\quad \vcx_1^2 = \vcx_2^2 = \vcx_3^2 = \vcx_4^2 = -1, \quad\quad \vcx_{\BA} \cdot \vcx_{\BB} = 0\quad(\BA\neq\BB).\enge

\noi Consider also  $\RR^{4,1}$, with basis $\{E_A\}_{A = 0}^4$, where 
\bege\label{r41} E_0^2 = -1, \quad\quad E_1^2 = E_2^2 = E_3^2 = E_4^2 = 1, \quad\quad E_A\cdot E_B = 0 \quad (A\neq B).
\enge
\noi The basis $\{E_A\}$ is obtained from the basis $\{\vcx_{\BA}\}$, if we define the isomorphism
\beq\label{pli}
\xi: \cl_{4,1} &\rightarrow& \la_2(\RR^{2,4})\nonumber\\
E_A &\mapsto& \xi(E_A) = \vcx_{A}\vcx_5.
\eeq
\noi The basis $\{E_A\}$ defined by eq.(\ref{pli}) obviously satisfies eqs.(\ref{r41}).

Given a vector $\ap = \ap^{\BA}\vcx_{\BA} \in \RR^{2,4}$, we obtain a paravector  $\mmb\in\RR\oplus\RR^{4,1}\hko\cl_{4,1}$ if the element $\vcx_5$ 
is left multiplied by $\mmb$:
\bege
 \mmb = \ap\vcx_5 = \ap^AE_A + \ap^5.
\enge

From the Periodicity Theorem, it follows the isomorphism $\cl_{4,1}\simeq\cl_{1,1}\otimes\clt$ 
and so it is possible to express an element of $\cl_{4,1}$ as a  $2\times 2$ matrix with entries in  $\cl_{3,0}$.

A homomorphism  $\vartheta: \cl_{4,1}\rightarrow\clt$ is defined as:
\beq
                 E_i &\mapsto& \vartheta(E_i) = E_iE_0E_4 \equiv \ee_i.
                    \eeq
                    It can be seen that  $\ee_i^2 = 1$, $E_i = \ee_iE_4E_0$ and 
                   
\bege
\bea{l} E_4 = E_+ + E_-,\\
        E_0 = E_+ - E_-,
        \ear
        \enge
where $E_\pm := \me(E_4 \pm E_0)$. Then,
\bege\label{mou}
\mmb = \ap^5 + (\ap^0 + \ap^4)E_+ + (\ap^4 - \ap^0) E_- + \ap^i\ee_i E_4E_0.
\enge
\noi  If we choose $E_4$ and $E_0$ to be represented by
\bege
E_4 =  \left(\bea{cc}
              0&1\\1&0\ear\right), \quad\quad  E_0 =  \left(\bea{cc}
              0&-1\\1&0\ear\right),\enge
consequently we have
\bege
E_+ =  \left(\bea{cc}
              0&0\\1&0\ear\right),\quad E_- =  \left(\bea{cc}
              0&1\\0&0\ear\right), \quad E_4E_0 =  \left(\bea{cc}
              1&0\\0&-1\ear\right),
              \enge
\noi and then the paravector $\mmb\in\RR\op\RR^{4,1}\hko\cl_{4,1}$ in eq.(\ref{mou}) is represented by
\bege\bea{|l|}\hline\\\mmb =  \left(\bea{cc}
              \ap^5 + \ap^i\ee_i&\ap^4 - \ap^0\\\ap^0 + \ap^4&\ap^5 - \ap^i\ee_i\ear\right)\\ \\ \hline\ear
              \enge                
              
 The vector $\ap\in\RR^{2,4}$ is in the Klein absolute, i.e., $\ap^2 = 0$. Besides, this condition implies that
 \bege \ap^2 = 0 \Leftrightarrow \mmb{\bar{\mmb}} = 0,\enge
 \noi since \beq
\ap^2 = \ap\ap &=& \ap 1\ap\n
&=& \ap\vcx_5^2\ap\n
 &=& \ap\vcx_5\vcx_5\ap\n
 &=& \mmb{\bar{\mmb}}.
\eeq 
We denote
 \bege\label{aloa}\bea{l}
 \lambda = \ap^4 - \ap^0,\\
 \mu = \ap^4 + \ap^0.\ear\enge Using the matrix representation of $\mmb{\bar{\mmb}}$, the entry
 ($\mmb{\bar{\mmb}})_{11}$ of the matrix is given by 
 \bege \label{xx}             
 (\mmb{\bar{\mmb}})_{11} = x{\bar{x}} -\lambda\mu = 0, 
 \enge
 \noi where 
\bege \label{xc}
x := (\ap^5 + \ap^i\ee_i)\in \RR\oplus\RR^3\hko\clt.
\enge\noi 
If we fix $\mu = 1$, consequently $\lambda = x{\bar{x}}\in \RR$. 
This choice does correspond to a projective description. Then the  paravector $\mmb\in\RR\oplus\RR^{4,1}\hko\cl_{4,1}$ can be represented as
 \bege\label{parax}
\mmb =  \left(\bea{cc}x &\lambda\\
 \mu&{\bar{x}}
 \ear\right) = \left(\bea{cc}x &x{\bar x}\\
 1&{\bar{x}}
 \ear\right).\enge
  From eq.(\ref{xx}) we obtain
  \beq
  (\ap^5 + \ap^i\ee_i)(\ap^5 - \ap^i\ee_i) = (\ap^4 - \ap^0)(\ap^4 + \ap^0) \nonumber\\
  \Rightarrow (\ap^5)^2  - (\ap^i  \ee_i)(\ap^j\ee_j) = (\ap^4)^2 - (\ap^0)^2,
  \eeq 
  \noi and conclude that 
  \bege\bea{|l|}\hline{\rule[-3mm]{0mm}{8mm} 
  (\ap^5)^2 + (\ap^0)^2 - (\ap^1)^2 - (\ap^2)^2 - (\ap^3)^2 - (\ap^4)^2 = 0}
  \\ \hline\ear\enge
  \noi which is the Klein absolute (eq.(\ref{klein})).

\subsection{ M\"obius transformations in Minkowski spacetime} 
  
The matrix $ g = \left(\bea{ll}a&c\\b&d\ear\right)$ is in the group \$pin$_+(2,4)$  if, and only if, its entries $a, b, c, d\in\cl_{3,0}$ 
satisfy the conditions \cite{maks}
\beq
(i)&&a{\bar{a}},\; b{\bar{b}}, \;c{\bar{c}},\; d{\bar{d}}\in\RR,\nonumber\\
(ii)&&a{\bar{b}},\; c{\bar{d}} \;\in\RR\oplus\RR^3,\nonumber\\
(iii)&&av{\bar{c}} + c{\bar{v}}{\bar{a}},\;\; cv{\bar{d}} + d{\bar{v}}{\bar{c}}\in \RR, \quad\forall v\in\RR\oplus\RR^{3},\nonumber\\
(iv)&&av{\bar{d}} + c{\bar{v}}{\bar{b}}\in \RR\oplus\RR^{3}, \quad\forall v\in \RR\oplus\RR^{3},\nonumber\\
(v)&&a{\tilde{c}} =  c{\tilde{a}},\; b{\tilde{d}} = d{\tilde{b}},\nonumber\\
(vi)&&a{\tilde{d}} - c{\tilde{b}} =   1.
\eeq
\noi Conditions $(i), (ii), (iii), (iv)$ are equivalent to the condition ${\hat{\si}}(g)(\mmb) := g\mmb {\tilde{g}}\in\RR\oplus\RR^{4,1}, \;\forall \mmb\in \RR\op\RR^{4,1}$, 
where ${\hat{\si}}: \${\rm pin}_+(2,4)\ri {\rm SO}_+(2,4)$ is the twisted adjoint representation. Indeed, 
\beq
g\mmb {\tilde{g}} &=&  \left(\bea{cc} a&c\\
 b&d
 \ear\right) \left(\bea{cc} x&\lambda\\
 \mu&{\bar{x}}
 \ear\right)\left(\bea{cc} {\bar d}&{\bar c}\\
 {\bar b}&{\bar a}
 \ear\right)\nonumber\\
  &=& \left(\bea{cc} ax{\bar{d}} + \lambda a{\bar{b}} + \mu c{\bar{d}} + c{\bar{x}}{\bar{b}}& ax{\bar{c}} + \lambda a {\bar{a}} + \mu c{\bar{c}} + bx{\bar{a}}\\
 bx{\bar{d}} + \lambda d{\bar{b}} + \mu d {\bar{d}} + d{\bar{x}}{\bar{b}}& bx{\bar{c}} + \lambda b{\bar{a}} + \mu d{\bar{b}} + d{\bar x}{\bar a}\ear\right)\nonumber\\
 &=&\left(\bea{ll} w&\lambda'\\ \mu'&{\bar{w}}\ear\right)\in\RR\op\RR^{4,1}
 \eeq\noi where the last equality  (considering $w\in\RR\oplus\RR^3$ and $\lambda',\mu'\in\RR$) comes from the requirement  that $g\in\${\rm pin}_+(2,4)$, i.e.,
  $g\mmb {\tilde{g}}\in\RR\oplus\RR^{4,1}$. If these conditions are required, $(i), (ii), (iii)$ and $(iv)$ follow.

 Conditions $(v), (vi)$ express $g{\bar{g}} = 1$, since for all $g\in\${\rm pin}_+(2,4)$, we have:  
\bege
g{\bar{g}} = 1\;\;\; \Leftrightarrow \;\;\;\left(\bea{ll}a{\tilde{d}} - c{\tilde{b}}&a{\tilde{c}} -  c{\tilde{a}}\\  
b{\tilde{d}} - d{\tilde{b}}&d{\tilde{a}} - b{\tilde{c}}\ear\right) = \left(\bea{ll}1&0\\0&1\ear\right).
\enge

  \subsection{Conformal transformations}

We have hust seen   that a paravector $\mmb\in\RR\op\RR^{4,1}\hko\cl_{4,1}$ is represented as 
 \bege
 \left(\bea{cc} x&x{\bar{x}}\\1&{\bar{x}}
 \ear\right) =  \left(\bea{cc} x&\lambda\\
 \mu&{\bar{x}}
 \ear\right),\enge
 \noi where $x\in\RR\op\RR^3$ is a paravector of $\clt$. 
 
 Consider an element of the group
\bege
 \${\rm pin}_+(2,4) := \{g \in \cl_{4,1} \;|\;g\bar{g} = 1\}.
\enge It is possible to represent it as an element  $g\in\cl_{4,1}\simeq\cl_{1,1}\otimes\clt$:
 \bege
  g = \left(\bea{cc} a&c\\
 b&d
 \ear\right),\quad a,b,c,d\in \clt. 
 \enge
 
 The rotation of  $\mmb\in\RR\op\RR^{4,1}\hko\cl_{4,1}$ is performed by the use of the twisted adjoint representation
 ${\hat{\si}}:\${\rm pin}_+(2,4) \ri {\rm SO}_+(2,4)$, defined as
\beq
{\hat{\si}}(g)(\mmb) &=& g\mmb{\hat{g}}^{-1}\n
 &=& g\mmb{\tilde{g}}, \qquad g\in\${\rm pin}_+(2,4).\eeq\noi
Using the matrix representation,  the action of \$pin$_+$(2,4) is given by
 \bege
 \left(\bea{cc} a&c\\
 b&d
 \ear\right) \left(\bea{cc} x&\lambda\\
 \mu&{\bar{x}}
 \ear\right){\widetilde{\left(\bea{cc} a&c\\
 b&d
 \ear\right)}} = \left(\bea{cc} a&c\\
 b&d
 \ear\right) \left(\bea{cc} x&\lambda\\
 \mu&{\bar{x}}
 \ear\right)\left(\bea{cc} {\bar d}&{\bar c}\\
 {\bar b}&{\bar a}
 \ear\right)
            \enge Fixing $\mu = 1$, the  paravector $\mmb$ is mapped on 
\bege
\left(\bea{cc} a&c\\
 b&d
 \ear\right) \left(\bea{cc} x&x{\bar{x}}\\
 1&{\bar{x}}
 \ear\right)\left(\bea{cc} {\bar d}&{\bar c}\\
 {\bar b}&{\bar a}
 \ear\right) = \Delta\left(\bea{cc} x'& x'{\bar{x'}}\\
 1&{\bar{x'}}
 \ear\right),
 \enge
\noi where \bege\label{acon} x' := (ax + c)(bx + d)^{-1},\enge \bege\Delta := (bx + d)({\ol{bx + d}})\in\RR.\enge 
The  transformation (\ref{acon}) is conformal \cite{vah,hes}. 

From the isomorphisms
\bege\label{isodirac}
\cl_{4,1}\simeq \CC\otimes\cle \simeq{\mt M}(4,\CC),
\enge\noi elements of \$pin$_+$(2,4) are elements of the Dirac algebra $\CC\otimes\cle$.  
  From eq.(\ref{xc}) we denote  $x\in\RR\oplus\RR^3$ a paravector. 
The conformal maps are expressed by the action of \$pin$_+$(2,4), by the following matrices:
\cite{port,maks,vah,hes}: 
\begin{center}
\begin{tabular}{||r||r||r||}\hline\hline
Conformal Map&Explicit Map&Matrix of $\$$pin$_+(2,4)$\\\hline\hline
Translation&$x\mapsto x + h,\;\; h\in \RR\op\RR^3$ &{\footnotesize{$\left(\bea{cc} 1&h\\
 0&1
 \ear\right)$}}\\\hline
Dilation&$x\mapsto \rho x, \;\rho\in\RR $& {\footnotesize{$
\left(\bea{cc} \sqrt{\rho}&0\\
 0&1/\sqrt{\rho}
 \ear\right)$}}\\\hline
Rotation&$x\mapsto \mmg x {\hat{\mmg}}^{-1},\; \mmg\in \${\rm pin}_+(1,3)$ &{\footnotesize{
$\left(\bea{cc} \mmg&0\\
0&{\hat{\mmg}}
 \ear\right)$}}\\\hline
Inversion&$x\mapsto -{\ol{x}}$& 
{\footnotesize{
$\left(\bea{cc} 0&-1\\
1&0
 \ear\right)$}}\\\hline
Transvection& $x\mapsto x + x(hx + 1)^{-1},\;\;h\in\ty$&  {\footnotesize{
$\left(\bea{cc} 1&0\\
h&1
 \ear\right) $}} \\\hline\hline
\end{tabular}                   
\end{center}
\noi This index-free geometric formulation allows to trivially generalize the conformal maps of $\RR^{1,3}$ to the ones of $\RR^{p,q}$,
 if the Periodicity Theorem of Clifford algebras is used. 

The group SConf$_+$(1,3) is fourfold covered by  SU(2,2), and the identity element $id_{{{\rm SConf}_+(1,3)}}$ of the group ${\rm SConf}_+(1,3)$
 corresponds to the following elements of SU(2,2)$\simeq$ \$pin$_+$(2,4):
\bege
\left(\bea{cc} 1_2&0\\
 0&1_2
 \ear\right),\quad\left(\bea{cc} -1_2&0\\
 0&-1_2
 \ear\right),\quad\left(\bea{cc} i_2&0\\
 0&i_2
 \ear\right),\quad\left(\bea{cc} -i_2&0\\
 0&-i_2
 \ear\right).
 \enge
 \noi  The element  $1_2$ denotes $id_{2\times 2}$ and $i_2$ denotes the matrix diag($i,i$).  

In this way, elements of \$pin$_+$(2,4) give rise to the orthochronous M\"obius transformations. The isomorphisms  
\bege
{\rm Conf}(1,3) \simeq {\rm O}(2,4)/\ZZ_2 \simeq {\rm Pin}(2,4)/\{\pm 1, \pm i\},
\enge
\noi are constructed in \cite{port} and consequently, 
\bege
{\rm SConf}_+(1,3) \simeq {\rm SO}_+(2,4)/\ZZ_2 \simeq {\rm \$pin}_+(2,4)/\{\pm 1, \pm i\}.
\enge 

The homomorphisms
\bege
\${\rm pin}_+(2,4) \stackrel{2-1}{\longrightarrow} {\rm SO}_+(2,4) \stackrel{2-1}{\longrightarrow} {\rm SConf}_+(1,3)\enge
\noi are explicitly constructed in \cite{Kl74,lau}.

\subsection{The Lie algebra of the associated groups}

 Consider $\cl^*_{p,q}$ the group of invertible elements. The function 
\beq
{\rm exp}:\cl_{p,q}&\ri&\cl^*_{p,q}\nonumber\\
a&\mapsto& {\rm exp}\;a = \sum_{n=0}^\infty \frac{a^n}{n!}
\eeq
\noi is defined. The vector space $V := \cl_{p,q}$ endowed with the Lie bracket is identified with the Lie algebra $\cl_{p,q}^*$. 

As an example, consider the  Clifford-Lipschitz $\Gamma_{p,q}$ group, a Lie subgroup of the Lie group
 $\cl^*_{p,q}$ and its Lie algebra is a vector subspace of $\cl_{p,q}$. Suppose that $X$
 is an element of $\Gamma_{p,q}$. Then ${\rm exp}(tX)$ is an element of $\Gamma_{p,q}$, i.e.,
\bege
f(t) = {\rm Ad\;exp}(tX)(\vv) \equiv {\rm exp}(tX)\;\vv\;{\rm exp}(-tX)\in \RR^{p,q}, \quad \forall\vv\in\RR^{p,q}.
\enge\noi Defining
\bege
({\rm ad}(X))(\vv) = [X,\vv] = X\vv - \vv X.
\enge\noi and using the well-known result
\bege
{\rm Ad}({\rm exp}(tX)) = {\rm exp}({\rm ad}(tX)),
\enge
\noi we have that  $f(t)\in\RR^{p,q}$ if, and only if
\bege
{\rm ad}(X)(\vv) = [X,\vv] = X\vv - \vv X \in \RR^{p,q}.
\enge
 It can be proved that  $X\in\Gamma_{p,q}$ is written as 
 \bege
 X \in {\rm Cen}(\cl_{p,q})\oplus\la_2(\RR^{p,q}).
 \enge
In this way, exp(t$X$)$\in \Gamma_{p,q}$. 
 
If $R\in{\rm Spin}(p,q)$, then ${\hat R} = R$, and for $R = {\rm exp}(tX)$, $X$ must be written as
 $X = a + B$, where $a\in\RR, B\in\la_2(\RR^{p,q})$. Besides, the condition  $R{\tilde{R}} = 1$ implies that
 $1 = {\rm exp}(t{\tilde{X}}){\rm exp}(tX) = {\rm exp}(2ta)$, i.e., $a = 0$. Then
 \bege\label{expalie}
\boxed{ {\rm Spin}_+(p,q)\ni R = {\rm exp}(tB), \quad B\in \la_2(\RR^{p,q})}
 \enge \noi 
 
The Lie algebra of Spin$_+(p,q)$, denoted by ${\mathfrak{spin}}_+(p,q)$ is generated by the space of 2-vectors endowed with the commutator. 
 Indeed, if $B$ and $C$ are bivectors, then
 \bege\label{bcz}
 BC = \langle BC\rangle_0 + \langle BC\rangle_2 + \langle BC\rangle_4,
 \enge\noi and since ${\tilde{B}} = -B$ e ${\tilde{C}} = -C$, it follows that
 \bege
 {\widetilde{BC}} = {\widetilde{C}}{\widetilde{B}} = CB.
 \enge
 \noi and so 
 \bege
\widetilde{BC} =  CB =  \langle BC\rangle_0 - \langle BC\rangle_2 + \langle BC\rangle_4,
 \enge\noi from where we obtain 
 \bege
 BC - CB = [B,C] = 2\langle BC\rangle_2.
 \enge
 \noi i.e., 
\bege
\boxed{(\la_2(\RR^{p,q}),\;[\;,\;]) = {\mathfrak{spin}}_+(p,q)}
\enge

\subsection{The Lie algebra of the conformal group}

The Lie algebra of \$pin$_+$(2,4) is generated by $\Lambda^2(\RR^{2,4})$, which has dimension 15. 
Since dim Conf(1,3) = 15, the relation between these groups is investigated now.
In the first paper of this series \cite{rol1}
we have just seen that
\bege \label{dre} E_0 = -i\g_0,\quad E_1 = -i\g_1,\quad E_2 = -i\g_2,\quad E_3 = -i\g_3,\quad E_4 = -i\g_{0123},\enge\noi and in the Sec. \ref{dcv}, that 
\bege\label{dri} E_A = \vcx_A\vcx_5,\enge \noi where $\{\vcx_{\BA}\}_{\BA = 0}^5$  is basis of $\RR^{2,4}$, $\{E_A\}_{A = 0}^4$ is basis of
 $\RR^{4,1}$ and $\{\g_\mu\}_{\mu = 0}^3$ is basis of $\RR^{1,3}$. 
The generators of Conf(1,3), as elements of $\la_2(\RR^{2,4})$, are defined as: 
\beq
P_\mu &=& \frac{i}{2}(\vcx_\mu\vcx_5 + \vcx_\mu\vcx_4),\nonumber\\
K_\mu &=& -\frac{i}{2}(\vcx_\mu\vcx_5 - \vcx_\mu\vcx_4),\nonumber\\
D &=& - \me \vcx_4\vcx_5,\nonumber\\
M_{\mu\nu} &=&  \frac{i}{2} \vcx_\nu\vcx_\mu.
\eeq 

From relations (\ref{dre}) and (\ref{dri}), the generators of Conf(1,3) are expressed from the  $\{\g_\mu\}\in\cle$ as
\beq
P_\mu &=& \me(\g_{\mu} + i\g_\mu\g_5),\nonumber\\
K_\mu &=& -\me (\g_{\mu} - i\g_\mu\g_5),\nonumber\\
D &=& \me i \g_5,\nonumber\\
M_{\mu\nu} &=& \me(\g_\nu\wedge\g_\mu).\nonumber\\
\eeq \noi They satisfy the following relations:
\beq
[P_\mu, P_\nu] &=& 0,\quad\quad [K_\mu, K_\nu] = 0,\quad\quad [M_{\mu\nu}, D] = 0,\nonumber\\  
 \left[M_{\mu\nu}, P_{\lambda}\right] &=& -(g_{\mu\lambda}P_\nu - g_{\nu\lambda}P_\mu),\nonumber\\ \left[M_{\mu\nu}, K_{\lambda}\right] &=& -(g_{\mu\lambda}K_\nu - g_{\nu\lambda}K_\mu),\nonumber\\  
 \left[M_{\mu\nu}, M_{\sigma\rho}\right] &=& g_{\mu\rho}M_{\nu\sigma} + g_{\nu\si}M_{\mu\rho} - g_{\mu\si}M_{\nu\rho} - g_{\nu\rho}M_{\mu\si},\nonumber\\ 
 \left[P_\mu, K_\nu\right] &=& 2(g_{\mu\nu} D - M_{\mu\nu}),\nonumber\\ 
 \left[P_\mu, D\right] &=& P_\mu,\nonumber\\
 \left[K_\mu, D\right] &=& -K_\mu.
\eeq \noi The commutation relations above are invariant under substitution $P_\mu \mapsto - K_\mu$, $K_\mu \mapsto - P_\mu$ and $D\mapsto -D$.

\section{Twistors as geometric  multivectorial elements}
In this section we present and discuss the Keller approach, and introduce our definition, showing how our twistor
formulation can be led to the Keller approach and consequently, to the Penrose classical twistor theory. The twistor
defined as a minimal lateral ideal is also given in \cite{Cw91,cru}.  Robinson
congruences and the incidence relation, that determines a spacetime point as  a secondary concept obtained from the
intersection between two twistors, are also investigated.

\subsection{The Keller approach}
\label{Keller}

The twistor approach by J. Keller \cite{ke97} uses the  projectors  $\PP_{\mt X} := \me(1 \pm i\g_5)$ (${\mt X} = {\mt R},{\mt  L}$) 
and the element  $T_{\xx} = 1 + \g_5{\xx}$, where ${\xx} = x^\mu\g_\mu\in\RR^{1,3}$. Now we introduce some results obtained by Keller \cite{ke97}: 
\medbreak
{\bf Definition} $\vvn$ The {\it reference twistor} $\eta_\xx$, associated with the vector $\xx\in\RR^{1,3}$ and a Weyl covariant dotted spinor (written as the 
left-handed projection of a Dirac spinor $\om$) $\Pi = \PP_{\mt L}\om = {0\choose \xi}$ is given by 
\bege\label{tr}\bea{|l|}\hline{\rule[-3mm]{0mm}{8mm}
\eta_\xx = T_\xx \PP_{\mt L}\om = (1 + \g_5\xx)\PP_{\mt L}\om = (1 + \g_5\xx)\Pi}\\ \hline\ear \vbn
\enge\noi  \medbreak 
In order to show the equivalence of this definition with the Penrose classical twistor formalism, the Weyl representation is used:
 \bege \eta_\xx = (1 + \g_5\xx)\Pi = \left[\left(\bea{cc}
           I&0\\
           0&I\ear\right) + \left(\bea{cc}
           -i_2&0\\
           0&i_2\ear\right)\left(\bea{cc}
           0&{\vec{x}}\\
           {\vec{x}}^c&0\ear\right)\right]{0\choose\xi}.\enge\noi Each entry of the matrices above denote  $2\times 2$ matrices, the vector 
           \bege\label{ref4}
           {\vec{x}} = \left(\bea{cc}
           x^0 + x^3 & x^1 + ix^2\\
           x^1 - ix^2 & x^0 - x^3\ear\right)\enge\noi is related to the point 
 $\xx\in\RR^{1,3}$ and ${\vec{x}}^c$ is the $\HH$-conjugation of  ${\vec{x}}\in\RR^{1,3}$ given by eq.(\ref{ref4}).
Therefore,
           \bege\label{fre}\bea{|l|}\hline{\rule[-4mm]{0mm}{10mm}
           \eta_\xx = \displaystyle{-i {\vec{x}}\xi\choose\xi}}\\ \hline\ear
           \enge\noi That is the index-free  version of Penrose classical  twistor \cite{pe1}. The sign in the first component is different, since it is used the Weyl representation. 
    \bege
\;\;\g(e_0) = \g_0 =  \left(\bea{cc}0&I\\
                                I&0\ear\right), \quad \g(e_k) = \g_k = \left(\bea{cc}
                                0&-\si_k\\
                                \si_k&0\ear\right).\enge\noi In order to get the correct sign, Keller uses a representation similar to the Weyl one,
but with the vectors in $\RR^3$ reflected (${\vec{x}}\mapsto - {\vec{x}}$) through the origin:
                                 \bege\;\;\g(e_0) = \g_0 =  \left(\bea{cc}0&I\\
                                I&0\ear\right), \quad \g(e_k) = \g_k = \left(\bea{cc}
                                0&\si_k\\
                                -\si_k&0\ear\right)\enge\noi Then it is possible to get the Penrose twistor
 \bege\label{fre1}\bea{|l|}\hline{\rule[-4mm]{0mm}{10mm}
           \eta_\xx = \displaystyle{i {\vec{x}}\xi\choose\xi}}\\ \hline\ear \enge

Twistors are completely described by the multivectorial structure of the Dirac algebra   $\CC\ot\cl_{1,3}\simeq\cl_{4,1}\simeq {\mt M}(4,\CC)$.

A classical spinor is an element that carries the irreducible representation of Spin$_+(p,q)$.  
Since this group is the set of even elements $\phi$ of the Clifford-Lipschitz  group such that 
 $\phi{\tilde{\phi}} = 1$, the irreducible representation comes from the irreducible representation of the even subalgebra $\cl_{p,q}^+$.  

\subsection{An alternative approach to twistors} 

 We now define twistors as a special class of algebraic spinors in $\cl_{4,1}$. 
The isomorphism  $\cl_{4,1}\simeq\CC\ot\cle$ presented in \cite{rol1} is given by 
\bege
E_0 = i\g_{0}, \quad E_1 = \g_{10},\quad E_2 = \g_{20}, \quad E_3 = \g_{30}, \quad E_4 = \g_5\g_0 = -\g_{123},
\enge
\noi and obviously $E_0^2 = -1$ and $E_1 = E_2 = E_3 = E_4 = 1$.
A paravector  $x\in\RR \op \RR^{4,1}\hko\cl_{4,1}$ is written as
\beq
 x &=& x^0 + x^AE_A\nonumber\\
 &=& x^0 + \ap^0E_0 + x^1E_1 + x^2E_2 + x^3E_3 + \ap^4E_4.
\eeq

We also define an element  $\chi := xE_4\in\la^0(\RR^{4,1})\op\la^1(\RR^{4,1})\op\la^2(\RR^{4,1})$ as
\beq
\chi &=& xE_4\n
 &=& x^0E_4 + \ap^0E_0E_4 + x^1E_1E_4 + x^2E_2E_4 + x^3E_3E_4 + \ap^4.
\eeq
\noi It can be seen that
\bege\label{fern}
\chi \me(1 + i\g_5) = T_\xx \me(1 + i\g_5) = T_\xx \PP_{\mt L}.
\enge\noi We define the twistor as: 
\bege\bea{|l|}\hline{\rule[-3mm]{0mm}{8mm}
{\rm Twistor} = \chi \PP_{\mt L}Uf \in (\CC\otimes\cle)f}\\ \hline\ear\enge\noi where $f$ is a primitive idempotent of
 $\CC\otimes\cle\simeq\cl_{4,1}$ and $U\in \cl_{4,1}$ is arbitrary. Therefore $Uf$ is a Dirac spinor and
 $ \PP_{\mt L}Uf = {0\choose \xi} = \Pi\in\me(1 + i\g_5)(\CC\ot\cl_{1,3})$ is a covariant dotted Weyl spinor. The twistor is written as
\beq
\chi \Pi &=& xE_4\Pi\nonumber\\
 &=& (x^0E_4 + \ap^0E_0E_4 + x^1E_1E_4 + x^2E_2E_4 + x^3E_3E_4 +\ap^4)\Pi.
\eeq
\noi Since
\beq
E_4\Pi &=& \g_5\g_0\Pi\n
 &=& -\g_0\g_5\Pi\n
 &=& -i\g_0\Pi.
\eeq
\noi  then it follows that
\beq
\chi \Pi &=&  (x^0E_4 + \ap^0E_0E_4 + x^1E_1E_4 + x^2E_2E_4 + x^3E_3E_4 + \ap^4)\Pi\nonumber\\
      &=&  x^0E_4\Pi + x^kE_kE_4\Pi + \ap^0 E_0E_4\Pi + \ap^4\Pi\nonumber\\
      &=&  x^0(E_4\Pi) + x^kE_k(E_4\Pi) + \ap^0 E_0(E_4\Pi) + \ap^4\Pi\nonumber\\
      &=&  x^0(-i\g_0\Pi) + x^k(\g_k\g_0)(-i\g_0\Pi) + \ap^0(i\g_0)(-i\g_0\Pi) + \ap^4\Pi\nonumber\\
      &=&  -ix^0\g_0\Pi - ix^k\g_k\Pi + \ap^0\Pi + \ap^4\Pi\nonumber\\
      &=&  -i{\xx} \Pi + \Pi\n  &=& (1 - i{\xx})\Pi\nonumber\\
      &=& (1 + \g_5\xx)\Pi\nonumber\\
      &=& \displaystyle{i {\vec{x}}\xi\choose\xi}.
     \eeq
\noi Then our definition is shown to be equivalent to the Keller one, and therefore, to the Penrose classical twistor, by eq.(\ref{fre1}).

 The incidence relation, that determines a point in spacetime from the intersection between two twistors \cite{pe3},  is given by
\beq
J_{{\bar{\chi}}\chi} &:=& {\ol{xE_4U}}xE_4U\n
 &=& -{\bar U}E_4 {\bar x}xE_4U\n
 &=& 0,\eeq \noi since the paravector $x\in \RR\oplus\RR^{4,1}\hko\cl_{4,1}$ is in the Klein absolute, and consequently, $x{\bar{x}} = 0$. 

Finally, the  Robinson congruence is defined in our formalism from the product
\beq
J_{{\bar{\chi}}\chi^\prime} &:=& {\ol{xE_4U}}x^\prime E_4U\n
 &=& -{\bar U}E_4 {\bar x}x^\prime E_4U.
\eeq \noi The above product is null if  $x = x'$ and the  Robinson congruence is defined when we fix $x$ and let  $x'$ vary.

\section*{Concluding Remarks}

The paravector model permits to express vectors of $\RR^{p,q}\hko\cl_{p,q}$ as paravectors, elements of $\RR\oplus\RR^{q,p-1}\hko\cl_{q,p-1}$. 
The conformal transformations (translations, inversions, rotations, transvections and dilations) are expressed via the adjoint representation of 
 \$pin$_+$(2,4) acting on paravectors of $\cl_{4,1}$. 
While the original formulation of the conformal transformations is described as rotations in $\RR^{2,4}$, 
the paravector model allows to describe them using the Clifford algebra $\cl_{4,1}$, isomorphic to the  Dirac-Clifford algebra $\CC\ot\cle$. 
Then the redundant dimension is eliminated. Also, the Lie algebra related to the conformal group is described via the Dirac-Clifford algebra.

Twistors are defined in the index-free Clifford formalism as particular algebraic spinors 
(with an explicit dependence of a given spacetime point) of $\RR^{4,1}$, i.e.,  twistors are elements of a left minimal ideal of the Dirac-Clifford algebra
  $\CC\otimes\cl_{1,3}$. Equivalently, twistors are classical spinors of $\RR^{2,4}$. 
Our formalism is led to the well-known formulations, e.g.,  Keller \cite{ke97} and Penrose \cite{pe3,pe4,pe5,pe1,pe2}. 
The first advantage of an index-free formalism is the explicit geometric nature of the theory, besides the more easy comprehension of an abstract index-destituted
theory.  
Besides, using the Periodicity Theorem of Clifford algebras, the present formalism can be generalized, 
in order to describe conformal maps and twistors in any ($2n$)-dimensional quadratic space. The relation between this formalism
and exceptional Lie algebras, and the use of the pure spinor formalism is investigated in \cite{eu}.

\end{document}